\documentclass[12pt]{article}

\usepackage{scicite}

\usepackage{times}
\usepackage{graphicx}  % needed for figures
\usepackage{epstopdf}
\usepackage{dcolumn}   % needed for some tables
\usepackage{bm}        % for math
\usepackage{amssymb}   % for math
\usepackage{amsmath}   % for matrix
\usepackage{amsthm}
\usepackage{chemarrow}
\usepackage{color}
\usepackage{mathrsfs}
\usepackage{float}
\usepackage{bbold}
\usepackage{bbm}
\usepackage[a4paper,colorlinks=true,
linkcolor=blue,citecolor=blue,
pdfauthor={ },
pdftitle={ },
pdfsubject={ },
pdfkeywords={ }]{hyperref}

\theoremstyle{plain}
\newtheorem*{theorem*}{Theorem}

\usepackage{lipsum} % µ÷ÓÿçÀ¸³¤¹«Ê½ºê°ü

\newenvironment{sciabstract}{%
\begin{quote} \bf}
{\end{quote}}

\begin{document}


\baselineskip24pt

% Include your paper's title here

\title{Floquet maser}

% Place the author information here.  Please hand-code the contact
% information and notecalls; do *not* use \footnote commands.  Let the
% author contact information appear immediately below the author names
% as shown.  We would also prefer that you don't change the type-size
% settings shown here.

\author
{Min Jiang$^{1,2}$, Haowen Su$^{1,2}$, Ze Wu$^{1,2}$, Xinhua Peng$^{1,2,3\ast}$, Dmitry Budker$^{4,5,6}$\\
\\
\normalsize{$^{1}$Hefei National Laboratory for Physical Sciences at the Microscale and Department of Modern Physics,}\\
\normalsize{University of Science and Technology of China, Hefei 230026, China}\\
\normalsize{$^{2}$CAS Key Laboratory of Microscale Magnetic Resonance,}\\
\normalsize{University of Science and Technology of China, Hefei 230026, China}\\
\normalsize{$^{3}$Synergetic Innovation Center of Quantum Information and Quantum Physics,}\\
\normalsize{University of Science and Technology of China, Hefei 230026, China}\\
\normalsize{$^{4}$Helmholtz-Institut, GSI Helmholtzzentrum f{\"u}r Schwerionenforschung, 55128 Mainz, Germany}\\
\normalsize{$^{5}$Johannes Gutenberg-Universit{\"a}t Mainz, 55128 Mainz, Germany}\\
\normalsize{$^{6}$Department of Physics, University of California, Berkeley, CA 94720-7300, USA}\\
\\
\normalsize{$^\ast$To whom correspondence should be addressed; E-mail: xhpeng@ustc.edu.cn}
}

% Include the date command, but leave its argument blank.

\date{}

\maketitle

\begin{sciabstract}
The invention of the maser stimulated many revolutionary technologies such as lasers and atomic clocks.
Despite enormous progress,
the realizations of masers are still confined to a limited variety of systems,
in particular, the physics of masers remains unexplored in periodically driven (Floquet) $\textrm{systems}$,
which are generally defined by time-periodic Hamiltonians and enable to observe many exotic phenomena such as time crystals.
Here we investigate the Floquet system of periodically driven $^{129}$Xe gas under damping feedback,
and surprisingly observe a multi-mode maser that oscillates at frequencies of transitions between Floquet states.
Our findings extend maser techniques to Floquet systems,
and open a new avenue to probe Floquet phenomena unaffected by decoherence,
enabling a new class of maser sensors.
As a first application,
our maser offers a unique capability of measuring low-frequency (1-100~mHz) magnetic fields with femtotesla-level sensitivity,
which is significantly better than state-of-the-art magnetometers,
and can be immediately applied to, for example, ultralight dark matter searches.
\end{sciabstract}

\noindent
\textbf{{\color{red}INTRODUCTION}}

\noindent
The masers\cite{Townes1955, Oxborrow2012, Breeze2018, Kraus2014} have become ubiquitous and resulted in innovations\cite{Maiman1960, Goldenberg1960,Cook1961,Chu2004, Suefke2017} ranging from
lasers, atomic clocks, ultrasensitive magnetic resonance spectroscopy, and low-noise amplifiers to deep-space communications.
Because the frequency of radiowaves produced by masers is highly stable,
these devices enable exquisitely sensitive measurement of their frequency shifts caused by the interactions with external electromagnetic fields.
This opens up exciting possibilities for developing precise metrology in applied and fundamental physics,
such as  magnetometry\cite{Gilles2003,Suefke2017, Bevington2020}, temperature sensors\cite{Jin2015}, tests of Lorentz and CPT violation\cite{Bear2000}, and searches for topological dark matter\cite{Derevianko2014}.
The masers have been demonstrated in a variety of systems,
such as ammonia molecules\cite{Townes1955}, hydrogen atoms\cite{Goldenberg1960}, noble gas\cite{Bear2000,Gilles2003,Asahi2000},
pentacene moleucles\cite{Oxborrow2012}, silicon and nitrogen vacancy defect materials\cite{Kraus2014,Breeze2018,Gilles2003}.
However, the demonstrations of $\textrm{masers}$ have remained unexplored in periodically driven (Floquet) systems\cite{Moessner2017},
limiting the broad applications in sensing, spectroscopy, and fundamental physics.
The generalization of masers to periodically driven systems would pave the way for many new applications,
such as ultralow-frequency magnetic field sensing\cite{Mateos2015, Marfaing2009} and
searching for oscillating electric dipole moments ($\textrm{EDMs}$) (see review in refs. \cite{DeMille2017, Safronova2018}).

To reach the above goal,
a proper periodically driven maser gain medium should be considered.
The recently developed notion of Floquet system (see review in ref. \cite{Moessner2017}),
which is only invariant under discrete time translations by a period,
has spawned intriguing prospects,
such as time crystals\cite{Zhang2017} and Floquet topological insulators\cite{Rechtsman2013}.
A variety of Floquet systems have been realized through periodic driving,
ranging from periodically driven trapped ions\cite{Zhang2017}, atomic ensembles\cite{Rechtsman2013,Eckardt2017} to nitrogen-vacancy centres\cite{Shu2018}.
For the goal of developing periodically driven masers,
Floquet systems should be an outstanding candidate.
We note that the combination of maser techniques and Floquet systems can overcome the decoherence effect and
thus permits a fresh look at many phenomena.
For example,
successful realization of maser using Floquet systems may open new opportunities
for observing long-range temporal dynamics\cite{Zhang2017, Rechtsman2013} and spectroscopy with sub-millihertz resolution,
%high-order sideband effect up to 1000 orders,
with important implications in quantum metrology\cite{Lang2015, Joas2017, Zhang1994}, for example,
for in searches for gravitational waves in eLISA with the bandwidth of 1-100~mHz\cite{Mateos2015},
measurement of the world-wide magnetic-background noise (including attempts at earthquake prediction)\cite{Marfaing2009},
and axion dark matter searches\cite{budker2014, Wu2019, Garcon2019}.
Despite these appealing features,
a demonstration of masers based on Floquet systems was heretofore lacking.

Here,
we report the first theoretical and experimental demonstration of a Floquet based maser comprised of periodically driven $^{129}$Xe spins in an vapor cell.
Unlike the common masers that exploit inherent transitions\cite{Townes1955, Oxborrow2012, Breeze2018, Kraus2014,  Maiman1960, Goldenberg1960,Cook1961},
our maser is based on the synthetic dimensions supported by Floquet states of the system.
We name the observed maser `Floquet maser',
which oscillates at the frequencies of transitions between Floquet states.
%The successful realization of maser based on Floquet system offers the potential for probing Floquet phenomena unaffected by decoherence.
Using our maser technique,
we indeed observe ultrahigh-resolution spectra of the Floquet system with a two orders of magnitude better resolution compared to that limited by decoherence.
As the spectral resolution is greatly increased,
a different regime emerges where high-order Floquet sidebands become significant and complex spectra are expected,
enabling accurate measurement of physical parameters,
e.g., atomic scalar and tensor polarizabilities\cite{Zhang1994}, magnetic fields\cite{Mateos2015}, ultralight bosonic exotic fields\cite{Garcon2019}, and multiphoton coherences\cite{Glasenapp2014}.
As a first application,
our maser constitutes a new quantum technology for measuring ultralow-frequency (1-100~mHz) magnetic fields with femtotesla-level sensitivity,
which is significantly better than state-of-the-art magnetometers\cite{Mateos2015, Greenberg1998, DBudker2007, Taylor2008}.
Moreover, we show that
the present maser technique allows us to achieve a search sensitivity for the coupling of axion dark matter to masing spins
well beyond the most stringent existing constraints\cite{Garcon2019, Wu2019}.

~\

\noindent
\textbf{{\color{red}RESULTS}}

\noindent
\textbf{Setup and Floquet system}

\begin{figure*}[t]
\centering
\includegraphics[width=0.9\columnwidth]{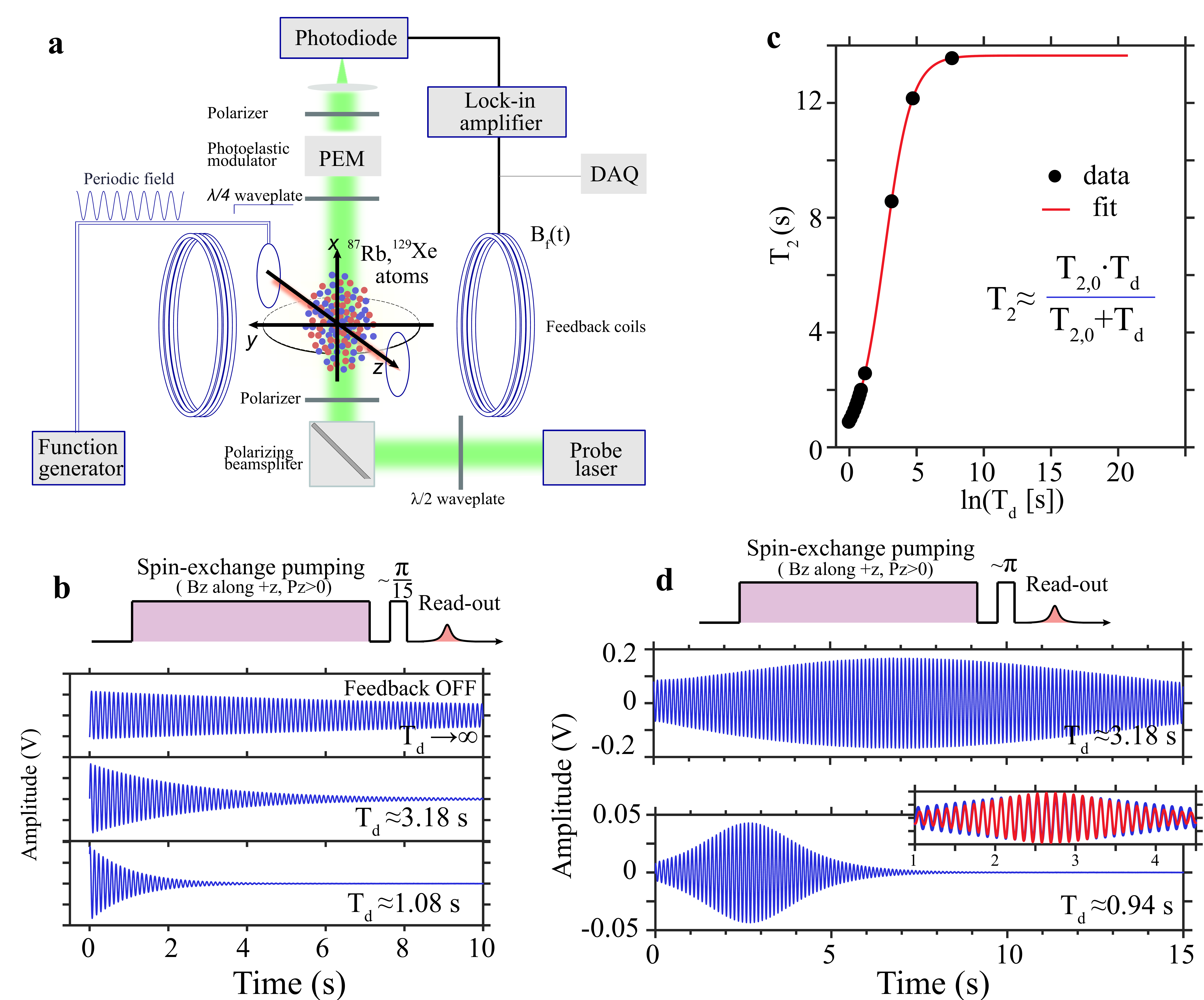}
    \caption{\textbf{Schematic of experimental setup and damping feedback mechanism}. $\textbf{a}$, Schematic of the Floquet $^{129}$Xe nuclear-spin maser. The $^{129}$Xe atoms are polarized and detected by spin-exchange collisions with optically pumped $^{87}$Rb (see Methods). Under a bias field and an oscillating magnetic field along ${z}$, the $^{129}$Xe spins are magnetically coupled to a feedback circuit, which feeds back real-time $B_{\textrm{f}}(t)$ along ${y}$ and induces the damping of $^{129}$Xe spins. $\textbf{b}$, Measured free decay $^{129}$Xe signals for three different feedback gains (corresponding to different $T_\textrm{d}$). Here the spin population is initialized to be positive (in this case, bias field is along $+z$). $T_\textrm{d}$ is well determined by corresponding decay time $T_2$ with $\sim \frac{\pi}{15}$ excitation angle (see main text). $\textbf{c}$, Measured decay time $T_2$ as a function of damping time (black symbols). The red line is a fit with $(T_{2,0}^{-1}+T_{\textrm{d}}^{-1})^{-1}$, where $T_{2,0}$ represents the intrinsic decoherence time without feedback. Here the spin population is initialized to be positive (in this case, bias field is along $+z$). $\textbf{d}$, Transient maser operations for two different damping times after flipping $\sim$$\pi$ angle, inducing the inversion of $^{129}$Xe spins population. The decay signal can be fitted with a hyperbolic secant function shown in the inset (see Methods).}
    \label{figure-1}
\end{figure*}

\noindent
We use noble gas atoms $^{129}$Xe with nuclear spin $I=1/2$ in a setup depicted in Fig.~\ref{figure-1}a.
A 0.5~cm$^3$ cubic vapor cell made from pyrex glass contains 5~torr $^{129}$Xe, 250~torr N$_2$, and a droplet of enriched $^{87}$Rb.
$^{129}$Xe spins are polarized by spin-exchange collisions with optically-pumped $^{87}$Rb atoms in a bias magnetic field $B_0$ ($\approx 750$~nT) along the polarized direction ($z$ axis)\cite{Walker1997}.
Similar to a microwave cavity in conventional masers\cite{Townes1955, Oxborrow2012, Breeze2018},
the $^{129}$Xe spins in our experiments are embedded in a feedback circuit\cite{Asahi2000} (see Fig.~\ref{figure-1}a),
which employs an atomic magnetometer\cite{DBudker2007} as a sensitive detector of $^{129}$Xe spins and
simultaneously supplies the real-time output audio-frequency signal of the magnetometer to the spins (see Methods).
The $^{87}$Rb atoms in the vapor cell act as a magnetometer for measuring the $^{129}$Xe spin polarization $P_x$ along the $x$ direction;
a corresponding feedback field $B_{\textrm{f}}(t)=\chi P_x(t)$ is applied to the $^{129}$Xe spins with a set of $y$ coils around the vapor cell.
Here the proportionality constant $\chi$ (feedback gain) encapsulates the conversion factor of the atomic magnetometer.
The feedback gain $\chi$ can be adjusted by a sliding rheostat in series with the feedback coils.
Similar to a resonant cavity\cite{Bienfait2016},
the self-induced feedback field $B_{\textrm{f}}(t)$ carries the information about the spins and then acts back on the spins,
leading to the well-known phenomenon of damping\cite{Bloembergen1954, Suefke2017} that is important in our maser scheme.

\begin{figure*}[t]
\centering
\includegraphics[width=0.9\columnwidth]{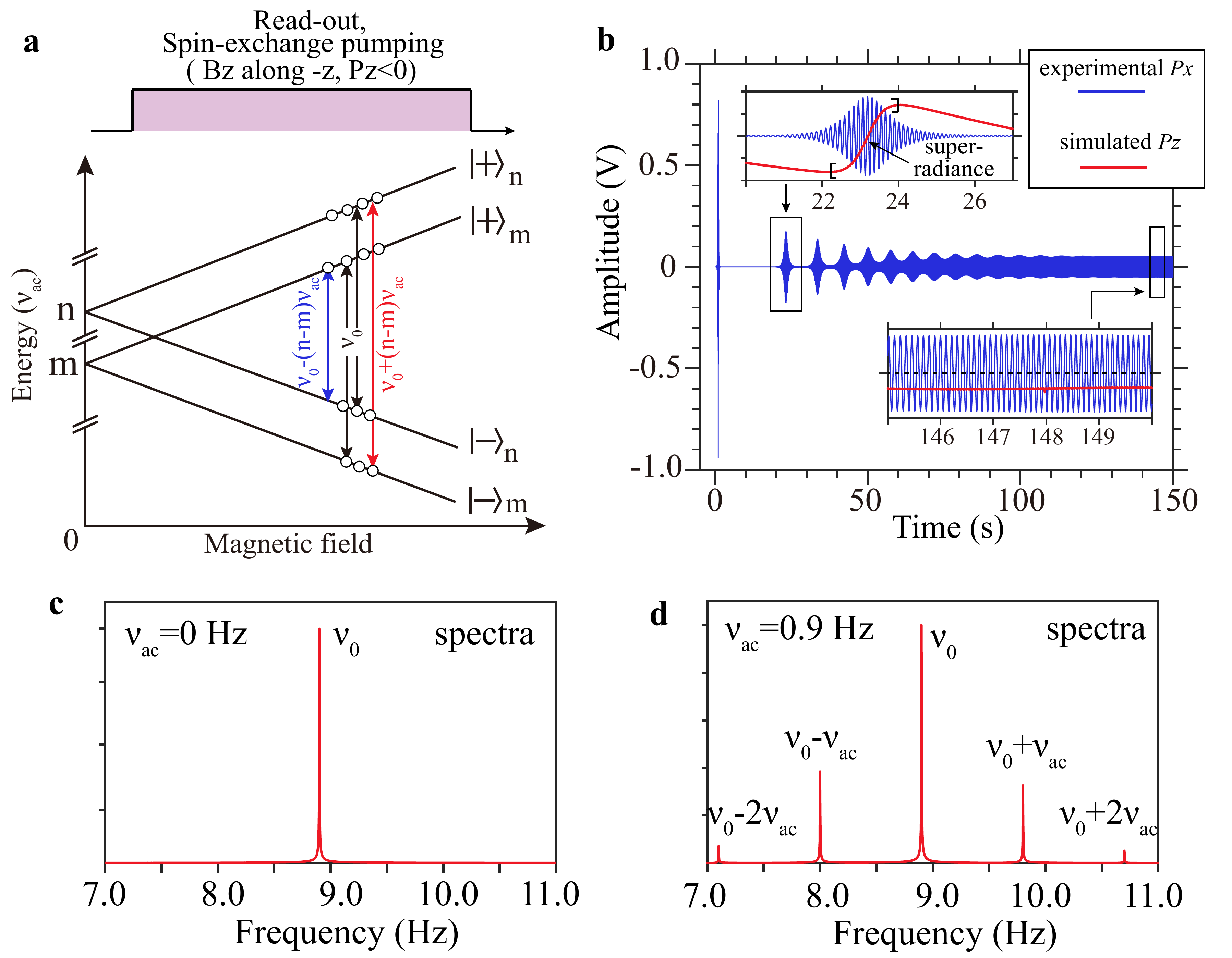}
    \caption{\textbf{Demonstration of Floquet maser}. $\textbf{a}$, Floquet states of a periodically driven two-level system (Floquet system). The energy gap between the upper and lower Floquet states $|+\rangle_n$ and $|-\rangle_m$ is $E_{n,m}/2\pi=(n-m)\nu_{\textrm{ac}} + \nu_0$. The spin population is initialized to be inverted (in this case, bias field is along $-z$). $\textbf{b}$, Signal of $^{129}$Xe Floquet maser. The insets are zoom-in plots for the signal and the simulated spin population ($P_{z}$). \textbf{c} and \textbf{d}, the corresponding amplitude spectra of the maser signal after eliminating the transient ($\nu_{\textrm{ac}}=0$ for \textbf{c}, 0.9~Hz for \textbf{d}).}
	\label{figure-2}
\end{figure*}

We consider a Floquet system,
where an oscillating magnetic field $B_{\textrm{ac}}\cos(2\pi \nu_{\textrm{ac}} t)$ (along $z$) periodically drives Zeeman energy levels.
Unfortunately, the textbook approach\cite{Townes1955, Oxborrow2012, Breeze2018, Kraus2014, Cook1961, Maiman1960, Goldenberg1960, Jin2015, Gilles2003} used to analyze masers is not well-suited for periodically driven systems.
%For example, an emitted-already photon passes back through the masing system, of which energy levels are time-varying, and thus cannot stimulate an identical photon.
As noted in the pioneering work\cite{Shirley1965},
a system with a time-periodic Hamiltonian can be equivalently represented as a time-independent Hamiltonian but with an infinite number of static energy levels.
Within this framework, we can borrow some essential notions from the conventional time-independent masers and this,
in turn, leads to the entirely new concept of Floquet maser.
Specifically,
% and the spin system becomes a Floquet system\cite{Moessner2017} with a period of $1/\nu_{\textrm{ac}}$.
a Floquet system has eigenstates (Floquet states) $|\pm\rangle_n=\sum_{n'} \mathcal{J}_{n-n'}(\pm \gamma B_{\textrm{ac}}/2\nu_{\textrm{ac}}) |\pm,n'\rangle$
and energies $E_{\pm,n}/2\pi=\pm \nu_0/2 + n\nu_{\textrm{ac}}$\cite{Shirley1965, Cohen-Tannoudji1992} (see Methods).
Here $|\pm,n'\rangle$ denotes that the spin is in the spin-up ($|+\rangle$) state or in the spin-down ($|-\rangle$) state
and the periodic driving field has the photon number $n'$\cite{Shirley1965, Cohen-Tannoudji1992}.
$\mathcal{J}_{n-n'}$ is the Bessel function of the first kind of order $n-n'$.
%Figure~\ref{figure-2}a shows the Floquet states, energies and transitions of a Floquet system.
Under periodic driving of the oscillating field,
the two-level ($|+\rangle$, $|-\rangle$) spin system is extended to an infinite number of synthetic energy levels $|\pm\rangle_n$ (see Fig.~\ref{figure-2}a) and
these energy levels are time-independent.

~\

\noindent
\textbf{Masing effect on Floquet system}

\noindent
We first measure the feedback-induced damping of $^{129}$Xe spins,
because damping plays an important role in realizing masers\cite{Bloembergen1954, Suefke2017, Asahi2000}.
%The radiation damping essentially represent a measure of the torque that is self-induced by the spin-feedback-circuit coupling.
For simplicity, the periodic driving field is turned off when we measure damping.
The measurement process is shown in Fig.~\ref{figure-1}b.
The $^{129}$Xe spins are initially polarized along $+z$ (bias field along $+z$),
and tilted by a small angle $\theta_0 \approx \frac{\pi}{15}$ along the $x$ axis,
then the free decay of $^{129}$Xe signals are measured under self-induced feedback.
In this case, the free decay signal can be fitted with a single-exponential decay
with a decay rate given by $T_{2,0}^{-1}+T_{\textrm{d}}^{-1}$~(see Methods),
where the intrinsic decoherence time $T_{2,0}\approx 13.65(1)$~s
and $T_{\textrm{d}}$ (the damping time) depends on feedback gain $\chi$.
By fitting the experimental data,
we can find the decay rate and then calculate the corresponding $T_{\textrm{d}}$ under different feedback gains $\chi$, as shown in Fig.~\ref{figure-1}b.
%It is worthy to finding that the radiation damping induces a frequency shift of the free-decay oscillation signal,
%which originates from Bloch-Siegert effect (\underline{see Supplementary Note 1}).
The results show that,
by coupling nuclear spins to the feedback circuit,
a regime can be reached in which damping constitutes the dominant mechanism of spin relaxation, e.g., $T_{\textrm{d}}=1.08(1)~ \textrm{s} \ll T_{2,0}$,
and spin relaxation can be controlled by adjusting the feedback gain (Fig.~\ref{figure-1}c).
This also suggests a method for active fast reset of long-lived spins (for example, $^3$He noble gas\cite{Gentile2017}) to their equilibrium state,
improving the repetition rate of an experiment.
%We show below that the controllable radiation damping is important for quantitatively studying maser.

We next measure the spin dynamics when the $^{129}$Xe spin population is suddenly inverted, corresponding to the case of $\theta_0 \approx  \pi$.
Figure~\ref{figure-1}d gives the observed free decay signals with designed $T_{\textrm{d}} \approx 3.18$~s and $0.94$~s, respectively.
Unlike the exponential decay,
the observed $^{129}$Xe spin signal first increases to a maximum value at a certain time and then decays to zero,
which can be described by a hyperbolic secant function (see Methods).
As first reported in ref.~\cite{Bloembergen1954},
this is a transient maser when the threshold of the damping time $T_{\textrm{d}}/ T_{2,0}\ll 1$ is fulfilled.
However, the demonstrated maser cannot oscillate continuously because the population inversion is transient.
In order to generate stationary maser dynamics\cite{Townes1955, Jin2015, Breeze2018},
we invert the direction of the bias magnetic field along $-z$ to prepare a continuous spin population (see Methods),
and simultaneously set the damping time smaller than the intrinsic decoherence time (i.e., $T_{\textrm{d}} / T_{2,0}< 1$).
Under these conditions, coupling of the spins to the damping feedback circuit can produce a self-sustained masing signal\cite{Asahi2000, Suefke2017}.

We now consider the spin dynamics of the Floquet $^{129}$Xe system under the damping feedback field.
As discussed above, Floquet system can be treated as a time-independent one with an infinite set of energy levels, shown in Fig.~\ref{figure-2}a.
%We introduce the notion of population inversion based on Floquet states.
The key to a maser based on Foquet systems is the preparation of spin population between those Floquet states.
In our experiments, population between Floquet states ($|+\rangle_n$ and $|-\rangle_m$) of the Floquet $^{129}$Xe spins can be continuously prepared
through $^{129}$Xe-$^{87}$Rb spin-exchange collisions when the bias field is set along $-z$.
Moreover, building on our demonstration of damping, the damping time is set to $T_{\textrm{d}}\approx 6.25$~s,
satisfying the threshold of $T_{\textrm{d}} / T_{2,0}< 1$.
When the feedback circuit is suddenly on,
a feedback $B_{\textrm{f}}(t)$ is induced by the Floquet system itself and oscillates with the frequencies of Floquet sidebands.
The feedback field produces a torque on the spins that changes spin polarization\cite{Asahi2000,Suefke2017}.
This self-coupling can lead to stimulated Rabi oscillations between the Floquet states $|+\rangle_n$ and $|-\rangle_m$.
A steady-state maser oscillation is expected to be build up.
For different Floquet states pair $n,m$, the maser oscillation frequency is $E_{n,m}/2\pi=(n-m)\nu_{\textrm{ac}} + \nu_0$.

As a first illustration of the Floquet maser,
Fig.~\ref{figure-2}b shows a time trace of $^{129}$Xe spins under 0.900-Hz driving field and $B_{\textrm{ac}}=56.15~\textrm{nT}$.
The $^{129}$Xe transverse polarization ($P_x$) shows characteristic initial transients,
which subsequently level into a stationary oscillation.
Because spin population ($P_z$) can not be measured directly,
we simulate the maser with nonlinear Bloch equations (see Methods) to gain important information on spin dynamics.
During the quick collapse of $P_x$ (see the top inset of Fig.~\ref{figure-2}b),
the negative spin population ($P_z<0$) reverses to positive ($P_z>0$) in a short time ($\sim 2$~s $\ll T_{1,0}\approx 21.5$~s),
this is the phenomenon of nuclear spin super-radiance\cite{Benedict2018}.
In the stationary window,
the spin population $P_z$ remains negative (see the bottom inset of Fig.~\ref{figure-2}b).
To extract the oscillation frequencies of the stationary maser,
we quantitatively investigate its spectrum.
We take a time trace from the stationary window after eliminating the initial transients,
and apply Fourier transform.
Unlike the common maser spectrum (without periodic driving) that has a single peak at $\nu_0 \approx 8.915$~Hz (Fig.~\ref{figure-2}c),
four evident sidebands appear at $\nu_{\pm 1}=\nu_0\pm 0.900$ and $\nu_{\pm 2}=\nu_0\pm 1.800$~Hz (Fig.~\ref{figure-2}d),
exactly at the frequencies of the transitions between Floquet states.
All lines are at regular intervals equal to the periodic driving frequency $\nu_{\textrm{ac}}$.
Our result confirms that a self-organized oscillation between Floquet states can indeed build up.
In addition, a 4000-s continuous oscillation gives a full-width at half-maximum (FWHM) of $0.3$~mHz,
which is two orders of magnitude narrower than the decoherence-limited resolution ($\approx 30$~mHz).

\begin{figure*}[t]
\centering
\includegraphics[width=0.6\columnwidth]{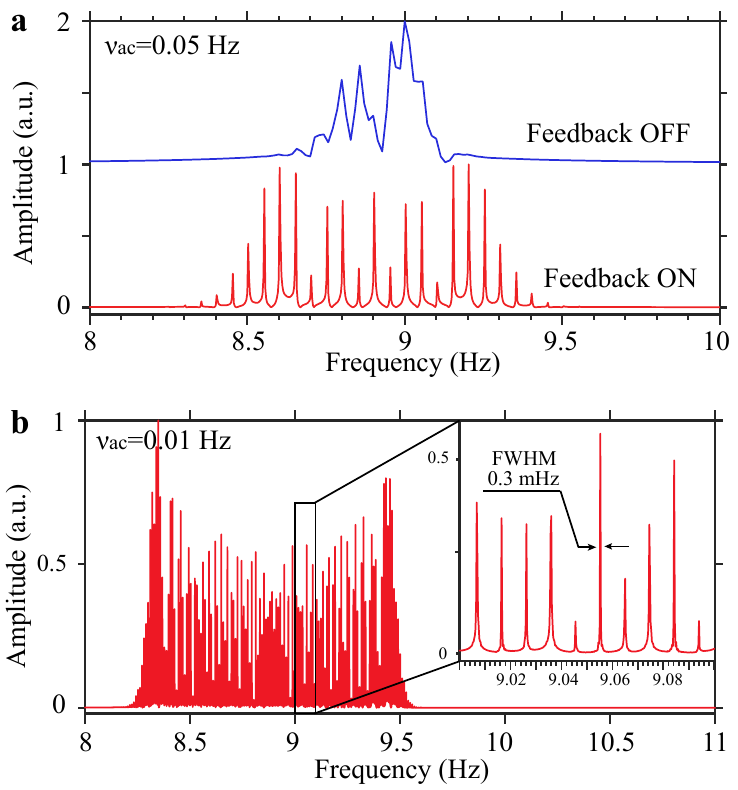}
    \caption{\textbf{Comb-like ultrahigh-resolution spectroscopy of Floquet maser}. $\textbf{a}$, Comparison between conventional Floquet spectrum based on $^{129}$Xe free-decay signal (blue lines) and a spectrum based on Floquet maser (red lines). The driving frequency is $\nu_{\textrm{ac}}=0.050$~Hz, and its amplitude is $B_{\textrm{ac}}=56.15$~nT. $\textbf{b}$, Spectrum based on Floquet maser, where $\nu_{\textrm{ac}}=0.010$~Hz and $B_{\textrm{ac}}=56.15$~nT.}
	\label{figure-3}
\end{figure*}

For a given Floquet system,
the coupling to the periodic driving is characterized by  the modulation index $\gamma B_{\textrm{ac}}/\nu_{\textrm{ac}}$.
The Floquet maser discussed above  were studied in the weak-coupling regime of $\gamma B_{\textrm{ac}}/\nu_{\textrm{ac}} \ll 1$.
Strong-coupling regime ($\gamma B_{\textrm{ac}}/\nu_{\textrm{ac}} \gg 1$) is indispensable in various applications of nonlinear atomic spectroscopy\cite{Glasenapp2014} and strong-field physics\cite{Fuchs2009}.
To reach this regime,
previous work usually required a large $B_\textrm{ac}$ and $\nu_{\textrm{ac}}$ larger than the decoherence-limited linewidth.
However,
a large magnetic field also affects the atomic magnetometer and deteriorates its performance\cite{Kominis2003}.
%thus limiting conventional $\gamma B_{\textrm{ac}}/\nu_{\textrm{ac}} \sim 1$.
We provide an alternative way to reach the strong-coupling regime by greatly decreasing the driving frequency $\nu_{\textrm{ac}}$ as low as sub-millihertz,
much below the decoherence-limited resolution.
Thus, the modulation index could be significantly larger than one.
For example,
if a 0.050~Hz field with magnitude of 56.15~nT drives $^{129}$Xe, $\gamma B_{\textrm{ac}}/\nu_{\textrm{ac}}\thickapprox 13.2$.
In Fig.~\ref{figure-3}a,
the sideband spectrum based on the Floquet maser (red line) exhibits at least 25 evident comb-like symmetric lines centered at the Larmor-frequency line (at $\nu_0$).
%and contrarily the spectrum based on the free-decay signal (blue line) shows indistinct lines.
Similarly,
for frequencies $\nu_{\textrm{ac}}$ lower than $T_{2,0}$-limited linewidth, e.g., $\nu_{\textrm{ac}}$$=$10~mHz with amplitude 56.15~nT,
corresponds to modulation index $\gamma B_{\textrm{ac}}/\nu_{\textrm{ac}}\thickapprox 66$.
In this case, we observe at least 134 maser sidebands (Fig.~\ref{figure-3}b),
obtained from 4000-s continuous oscillation.
Compared with the maser in the weak-coupling regime,
a great number of Floquet transitions can build up stationary maser oscillations in the strong-coupling regime,
enabling to exact reconstruction of the Floquet energy levels.

We emphasize the difference between our maser and existing masers.
First, our maser is based on transitions between Floquet states,
whereas existing masers usually exploit inherent transitions\cite{Townes1955, Oxborrow2012, Breeze2018, Kraus2014, Cook1961, Maiman1960, Goldenberg1960, Jin2015, Gilles2003}.
Our maser generates sidebands that are easily tunable by changing the frequency and amplitude of the periodic driving.
As shown in this work,
the Floquet maser is well-suited for sensing oscillating driving field with a frequency resolution,
independent of the masing spin decoherence and limited only by the stability of the maser.
Second,
unlike the conventional maser that uses a microwave cavity\cite{Townes1955, Oxborrow2012, Goldenberg1960, Jin2015, Breeze2018},
our maser makes use of a feedback circuit based on the signal of an atomic magnetometer to provide damping feedback,
enabling the maser frequency down to the audio-frequency range.
Even though the oscillation frequency is much smaller than that of earlier established microwave masers\cite{Townes1955, Oxborrow2012, Goldenberg1960, Jin2015, Breeze2018},
we show below that our demonstrated maser could be particularly useful for precision measurements,
such as ultralow-frequency magnetic field sensing\cite{Mateos2015, Marfaing2009} and searching for ultralight new particles\cite{budker2014, Wu2019, Garcon2019}.

~\

\noindent
\textbf{Applications of Floquet maser in magnetometry}

\noindent
Recently, experimental investigations have been reported towards achieving high sensitivity of measuring magnetic fields in low-frequency regime (1-100 mHz),
which is of importance in applied\cite{Marfaing2009} and fundamental physics\cite{Mateos2015, budker2014, Wu2019, Garcon2019}.
However, it still remains challenging for state-of-the-art magnetometers to reach femtotesla-level sensitivity owing to significant $1/f$ noise.
In this work, we combine the demonstrated Floquet maser technique with magnetometry
and realize a sensitive magnetometer that is well suited to operating in the ultralow-frequency regime.
The idea of our magnetometer is that the measured oscillating field applied to the spins can be seen as a periodic driving that generates sidebands around the maser oscillation frequency ($\nu_0\gg 1$~Hz).
Thus, the maser realizes up-converts of the low-frequency field to a higher frequency, where the $1/f$ noise is negligible.
As we show below, our maser-based magnetometer can achieve femtotesla-level sensitivity in the millihertz range.

\begin{figure*}[t]
\centering
\includegraphics[width=0.6\columnwidth]{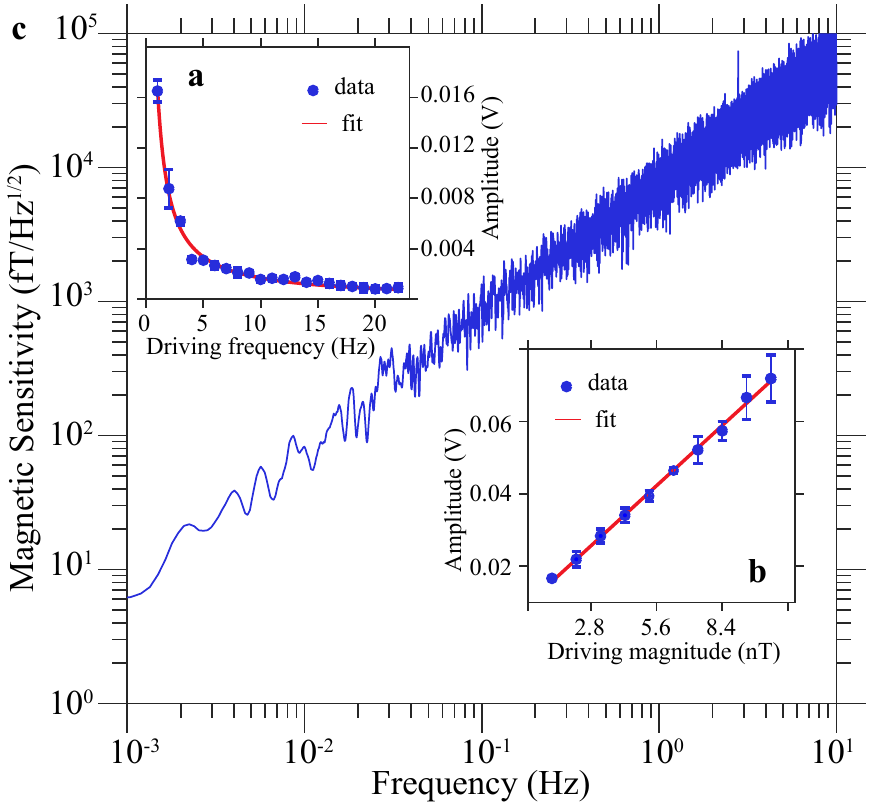}
    \caption{\textbf{Maser-based magnetometry on the first-order Floquet sideband of $^{129}$Xe}. $\textbf{a}$, The first-order Floquet sideband amplitude follows a $1/\nu_{\textrm{ac}}$ dependence with the driving frequency $\nu_{\textrm{ac}}$. The result of fit is $\xi=0.017/\nu_{\textrm{ac}}$; $B_{\textrm{ac}}=2.25$~nT. $\textbf{b}$, The first-order Floquet sideband amplitude follows a linear dependence with the driving amplitude $B_{\textrm{ac}}$. The fit result is $\xi=0.0055 B_{\textrm{ac}}+0.0096$, where $\nu_{\textrm{ac}}=1.000$~Hz. $\textbf{c}$, Measured magnetic sensitivity of the maser-based magnetometer (note that the vertical and horizontal axes have a logarithmic scale).}
	\label{figure-4}
\end{figure*}

We first calibrate the magnetic response of the maser by applying an oscillating magnetic field with known amplitude and frequency.
We focus on the prominent first-order sidebands occurring at $\pm \nu_{\textrm{ac}}$ about the central frequency,
whose amplitude is proportional to $\mathcal{J}_1(\gamma B_{\textrm{ac}}/\nu_{\textrm{ac}})\approx \gamma B_{\textrm{ac}}/2\nu_{\textrm{ac}}$
(here we assume that the oscillating fields are small enough, satisfying $\gamma B_{\textrm{ac}}/\nu_{\textrm{ac}} \ll 1$).
As shown in Fig.~\ref{figure-4}a, a 2.25-nT magnetic field is applied along the ${z}$ direction, and its frequency $\nu_{\textrm{ac}}$ is swept from 1~Hz to 22~Hz.
We Fourier-transform the individual measurement traces (with 60-s acquisition time) and record the corresponding sideband-peak amplitudes.
The experimental amplitudes of first-order sidebands are fitted to a $1/\nu_{\textrm{ac}}$ function,
which is in agreement with theory.
%The noise floor is shown in the subset of Fig.~\ref{figure4}(a).
Similarly, we measure the amplitude response by applying a $\nu_{\textrm{ac}}=1.000$~Hz magnetic field with varying amplitude, as shown in Fig.~\ref{figure-4}b.
The response of the maser to the applied oscillating field is measured to be $\xi_{\textrm{maser}} [\textrm{V}] \approx 5.5\cdot 10^{-3} B_{\textrm{ac}} [\textrm{nT}]/\nu_{\textrm{ac}} [\textrm{Hz}]$.

To show the noise performance of the maser-based magnetometer,
we take a 4000-s maser real-time data without periodic driving, and evaluate the corresponding power spectral density outside the spectral peak at $\nu_0$.
The background noise is white in the frequency range of maser sidebands,
and is measured to be $\delta \xi_{\textrm{maser}} \approx 4 \times 10^{-5}~\textrm{V}/\textrm{Hz}^{1/2}$.
Combining with the above-calibrated response,
the magnetic field sensitivity of maser-based magnetometer is estimated to be $\delta B_{\textrm{ac}} \approx  7.2 \nu_{\textrm{ac}} ~\textrm{pT}/\textrm{Hz}^{1/2}$ (Fig.~\ref{figure-4}c).
For example, our maser-based magnetometer has shown a sensitivity of about 7.2~$\textrm{fT}/\textrm{Hz}^{1/2}$ at 1~mHz,
which is significantly better than that in earlier work\cite{Mateos2015, Greenberg1998, DBudker2007, Taylor2008}.
Moreover, our result shows different frequency dependence: in our case, instead of the usual $1/\nu_{ac}$, the sensitivity scales as $\delta B_{\textrm{ac}} \propto  \nu_{\textrm{ac}}$.
For example, the magnetic sensitivity at $1$~mHz is 1000-fold better than that at $1$~Hz.
At the moment, for frequencies below 1~mHz,
the sensitivity of our maser deteriorates due to the maser instability.
The present device, therefore, is complementary to the state-of-the-art SQUIDs\cite{Greenberg1998} and atomic\cite{DBudker2007} magnetometers that have high sensitivity above 1~Hz.
Our device is particularly sensitive between 1 mHz and 1 Hz.
Moreover,
our magnetometer can operate in a nonzero magnetic field (e.g., in Earth's field),
in contrast with SERF atomic magnetometers that operate at fields below $\sim$100 nT\cite{Kominis2003}.

~\

\noindent
\textbf{{\color{red}DISCUSSION}}

\noindent
We have reported a novel maser based on time-periodic Floquet systems.
Unlike conventional masers, our maser oscillates at frequencies of transitions between Floquet states.
%Here, we report a novel maser based on the synthetic dimensions supported by Floquet states of Floquet $^{129}$Xe spins.
The generalization of the notion maser to periodically driven systems opens a new avenue to explore Floquet physics unaffected by decoherence effects.
% and develop a new class of maser sensors.
%in particular for exploring the spectroscopy and time-domain dynamics of Floquet systems without the intrinsic decoherence effect.
As we show in this work,
the connection of maser technique and Floquet physics allows observing ultrahigh-resolution spectra of Floquet systems with sub-millihertz widths
as well as high-order sidebands effect.
This can greatly improve the accuracy of measuring system energies, magnetic fields, atomic scalar and tensor polarizabilities\cite{Zhang1994},
ultralight bosonic exotic fields\cite{Garcon2019}, nonlinear multiphoton coherences\cite{Glasenapp2014}, etc.
Although demonstrated for $^{129}$Xe spins,
our scheme of Floquet maser can be transferred to other types of experimental systems.
For example, recent advances in cold atoms\cite{Rechtsman2013,Eckardt2017} and dipolar spin ensembles\cite{Shu2018} have led to progress in
such areas of Floquet physics as time crystals
and masers (for example, diamond maser).
We suggest future theoretical and experimental studies of masers based on a variety of Floquet systems,
permitting a fresh look at many phenomena, such as
Floquet Raman transitions\cite{Shu2018},
Mollow-triplet sidebands\cite{Joas2017},
Autler-Townes splitting\cite{Tannoudji1996},
and even time crystals\cite{Zhang2017, Rechtsman2013, Shu2018}.

Our Floquet technique is generic and can be easily applied to well-established masers,
such as hydrogen masers\cite{Goldenberg1960}, diamond masers\cite{Jin2015, Breeze2018}, and one-atom Rydberg maser\cite{Meschede1985},
all promising as a new class of maser-based quantum sensors.
For example, the application to hydrogen and diamond masers,
whose gyromagnetic ratios are about three orders of magnitude larger than those of nuclei,
yields improvements over currently achievable magnetic field sensitivity.
As we show in this work,
such sensors outperform state-of-the-art magnetometers\cite{Mateos2015, Greenberg1998, DBudker2007, Taylor2008} with femtotesla-sensitivity at millihertz frequencies,
and can be immediately applied, for example, in the searches of gravitational wave observatory eLISA (1-100~mHz)\cite{Mateos2015}.
Moreover,
we show for the first time that couplings between masing spins (such as protons or electrons) and
oscillating exotic fields beyond the standard model\cite{DeMille2017} may enable their direct detection via Floquet-maser magnetometry:
the exotic fields are predicted to couple with standard-model particles (masing spins in this case) and behave as an oscillating magnetic field,
generating sidebands\cite{Garcon2019} on masing spins that can be measured with our maser.
The maser can be applied to search for some exotic fields, such as ultralight axions\cite{budker2014, Wu2019, Garcon2019} and other exotic spin-dependent interactions\cite{DeMille2017}.
Compared with existing approaches, our work has a unique advantage of measuring exotic fields with ultralow frequency 1-100~mHz
(corresponding to particle mass $\sim10^{-18}$ to $10^{-15}$~eV).
As discussed in Methods, our work is promising for improving the search sensitivity of axions by approximately four and
five orders of magnitude compared to the values obtained in recent ultralow-field NMR experiments\cite{Wu2019, Garcon2019}.

~\

\noindent
\textbf{{\color{red}MATERIALS AND METHODS}}

\noindent
\textbf{Experimental apparatus}

\noindent
A cubic vapor cell containing 5~torr $^{129}$Xe, 250~torr N$_2$-buffer gas, and a droplet of enriched $^{87}$Rb
is placed inside a five-layer cylindrical mu-metal shield,
and is resistively heated to $140$~$^\circ$C.
The longitudinal and transverse relaxation times of $^{129}$Xe spins are measured to be $T_{1,0} \approx 21.5(2)$~s and $T_{2,0} \approx 13.65(1)$~s in a $z$ bias field ($\approx 750$~nT).
As shown in Fig.~\ref{figure-1}a,
$^{129}$Xe atoms are polarized by spin-exchange collisions with $^{87}$Rb atoms\cite{Walker1997},
which are pumped with circularly polarized laser light ($\approx$10 mW) propagating along $+z$.
The pump laser frequency is tuned to the center of the buffer-gas broadened and shifted D1 line.
The $^{87}$Rb atoms also act as a sensitive magnetometer for measuring the nuclear magnetization of the $^{129}$Xe spins
via optical rotation of linearly polarized probe laser light ($\approx1$~mW) propagating along $x$.
The frequency of the probe laser is detuned from the D2 transition by about 100 GHz.
In the presence of a $z$-bias field,
the $^{87}$Rb magnetometer is primarily sensitive to the $^{129}$Xe $x$-magnetization producing an effective magnetic field $B_{\textrm{eff}}=\frac{8\pi}{3}\kappa_0 M_x$\cite{Walker1997},
%The effective magnetic field of $^{129}$Xe experienced by the $^{87}$Rb atoms is given by $B_{\textrm{eff}}=\frac{8\pi}{3}\kappa_0 M_y$,
where the enhancement factor $\kappa_0$ is about 500,
and $x$-magnetization is $M_x=\mu_{\textrm{Xe}} n_{\textrm{Xe}} P_x$ with $\mu_{\textrm{Xe}}$, $n_{\textrm{Xe}}$, $P_x$
being the $^{129}$Xe nuclear magnetic moment, atomic density, and $x$-polarization, respectively.
To suppress the influence of low-frequency noise,
the polarization of the probe laser beam is modulated at 50~kHz with a photoelastic modulator (PEM),
and the signal is demodulated with a lock-in amplifier.

~\

\noindent
\textbf{Energies and states of Floquet systems}

\noindent
A two-level spin system driven with a radio-frequency (rf) field can be treated as a dressed-spin system\cite{Cohen-Tannoudji1992, Shirley1965},
in which rf photons with creation and annihilation operators $\hat{a}^\dag$ and $\hat{a}$ are introduced by the second quantization of the rf field.
Here, the rf field $B_{\textrm{ac}} \textrm{cos}(2\pi \nu_{\textrm{ac}}t)$
is applied parallel to the static magnetic field $B_0 \hat{z}$.
The second-quantized Hamiltonian for the dressed-spin system can be written as\cite{Cohen-Tannoudji1992}
\begin{equation}
\hat{H}/2\pi= \nu_{0} I_z + \nu_{\textrm{ac}} \hat{a}^\dag \hat{a}+ \lambda I_z (\hat{a}^\dag+ \hat{a}),
\label{H}
\end{equation}
where $\nu_0=\gamma B_0$.
The first term in Eq.~(\ref{H}) is the Zeeman interaction of the spin with $B_0$. %, and $\nu_0=\gamma B_0$ with the spin's gyromagnetic ratio $\gamma$.
The second term is the energy of the quantized rf field.
The final term describes the coupling between the spin and the quantized rf field with strength $\lambda=\gamma B_{\textrm{ac}}/(2\sqrt{\bar{N}})$\cite{Cohen-Tannoudji1992},
where $\bar{N} \gg 1$ is the average number of rf photons in the mode.
We introduce basis states $|\pm, n\rangle$,
where $|\pm,n\rangle=|\pm \rangle\otimes |n\rangle$, $n$ signifies the rf photon number of the driving field,
and $|\pm \rangle$ denotes the eigenstate of $\sigma_z$, i.e., $|+\rangle=(1,0)^\textrm{T}$ and $|-\rangle=(0,1)^\textrm{T}$.
The Hamiltonian $\hat{H}$ commutes with $\sigma_z$ and its diagonalization is reduced to that of $H_{\epsilon}$ onto each of the two eigensubspaces of $\sigma_z$,
\begin{equation}
\begin{array}{cccc}
\hat{H}_{\epsilon}/2\pi= \frac{\epsilon}{2} \nu_{0} + \nu_{\textrm{ac}} \hat{a}^\dag \hat{a}+ \frac{\epsilon}{2} \lambda  (\hat{a}^\dag+ \hat{a}),
\end{array}
\label{H1}
\end{equation}
where $\epsilon=\pm 1$.
A displacement operator is defined as $\textrm{D}(\xi)=e^{\xi \hat{a}^\dagger -\xi^{\ast} \hat{a}}$,
which has the following properties:
$\textrm{D}(\xi)\textrm{D}^{\dagger}(\xi)=1$,
$\textrm{D}(\xi) \hat{a}^\dagger \textrm{D}^{\dagger}(\xi)=\hat{a}^{\dagger}-\xi$,
and $\textrm{D}(\xi) \hat{a} \textrm{D}^{\dagger}(\xi)=\hat{a}-\xi$.
Then $\hat{H}_{\epsilon}$ in Eq.~(\ref{H1}) can be written as follows
\begin{equation}
\begin{array}{cccc}
\hat{H}_{\epsilon}/2\pi=\textrm{D}(-\frac{\epsilon \lambda}{2\nu_{\textrm{ac}}}) ( \frac{\epsilon}{2} \nu_{0} + \nu_{\textrm{ac}} \hat{a}^\dag \hat{a}-\frac{\lambda^2}{4\nu_{\textrm{ac}}}) \textrm{D}^{\dagger}(-\frac{\epsilon \lambda}{2\nu_{\textrm{ac}}}).
\end{array}
\label{H2}
\end{equation}
This Hamiltonian $\hat{H}_{\epsilon}$ has eigenstates
$|\epsilon\rangle_n=\textrm{D}^{\dagger}(-\frac{\epsilon \lambda}{2\nu_{\textrm{ac}}})|\epsilon, n\rangle$, i.e., the Floquet states.
The energy of $|\epsilon\rangle_n$ is $E_{\epsilon,n}/2\pi=\epsilon \nu_0/2 + n\nu_{\textrm{ac}}$.
%where $|\epsilon,n\rangle=|\epsilon\rangle\otimes |n\rangle$, and $|\epsilon\rangle$ denotes the eigenstate of $\sigma_z$, i.e., $|+\rangle=(1,0)^\textrm{T}$ and $|-\rangle=(0,1)^\textrm{T}$.
We now derive the explicit form of $|\epsilon\rangle_n$ for $n\approx \bar{N}\gg 1$ in the basis of $|\epsilon, n\rangle$.
Let $\xi=-\frac{\epsilon \lambda}{2\nu_{\textrm{ac}}}$.
By using the Glauber formula: $e^{A+B}=e^A e^B e^{[A,B]/2}$, where the two operators $A$ and $B$ both commute with their commutator,
we have
%\begin{widetext}
\begin{equation}
\begin{array}{cccc}
\langle n-m|e^{\xi \hat{a}^\dagger -\xi^{\ast} \hat{a}}|n\rangle \approx e^{-\frac{\xi^{\ast}\xi}{2}} \mathcal{J}_{-m} (2\xi \sqrt{\bar{N}}),
\end{array}
\label{H4}
\end{equation}
%\end{widetext}
%where the values of $p,q$ contributing to Eq.~(\ref{H4}) are on the order of $p,q \sim \lambda \sqrt{n}/\nu_{\textrm{ac}}\sim \gamma B_{\textrm{ac}}/\nu_{\textrm{ac}}\ll \bar{N}$.
%Thus, in the calculation of $\hat{a}^q |n\rangle$, we can take $n-p\approx n \approx \bar{N}$, which leads to
%$\hat{a}^q |n\rangle\approx \sqrt{\bar{N}^q}|n-q\rangle$.
As $|\xi|=\frac{ \lambda}{2\nu_{\textrm{ac}}}=\frac{\gamma B_{\textrm{ac}}}{\nu_{\textrm{ac}}} \frac{1} {4\sqrt{\bar{N}}} \ll 1$, $e^{-\frac{\xi^{\ast}\xi}{2}}\approx1$.
Also $2\xi \sqrt{\bar{N}}=-\frac{\epsilon\gamma B_{\textrm{ac}}}{2\nu_{\textrm{ac}}}$.
Finally, we obtain
$\langle n-m|e^{\xi \hat{a}^\dagger -\xi^{\ast} \hat{a}}|n\rangle=\mathcal{J}_{m} (\frac{\epsilon\gamma B_{\textrm{ac}}}{2\nu_{\textrm{ac}}})$,
and the Floquet states are
\begin{equation}
\begin{array}{cccc}
|\epsilon\rangle_n &=&|\epsilon\rangle \sum_m |n-m\rangle \langle n-m|e^{\xi \hat{a}^\dagger -\xi^{\ast} \hat{a}}|n\rangle\\
&=& \sum_{n'} \mathcal{J}_{n-n'} (\frac{\epsilon \gamma B_{\textrm{ac}}}{2\nu_{\textrm{ac}}}) |\epsilon, n'\rangle.
\end{array}
\label{H5}
\end{equation}
As a result of the periodic driving,
the two-level ($|+\rangle$, $|-\rangle$) spin system is extended to an infinite number of synthetic energy levels $|\epsilon \rangle_n$,
as shown in Fig.~\ref{figure-2}a.

~\

\noindent
\textbf{Maser mechanism with damping feedback}
%This section presents a numerical simulation of the Floquet-state maser.

\noindent
The feedback circuit employs a rubidium magnetometer to measure the nuclear polarization that acts back on the spins.
%The radiation damping on the nuclear spins embedded in the feedback circuit influences the dynamics of spins.
The dynamics of the $^{129}$Xe spin polarization $\vec{\mathbf{P}}=[P_x, P_y, P_z]$ is described by the nonlinear Bloch equations\cite{Suefke2017, Asahi2000},
%\begin{widetext}
%\begin{eqnarray}
\begin{equation}
\begin{cases}
\frac{d P_x}{dt}= P_y \omega_z (t) - P_z \omega_y (t) -\frac{1}{T_{2,0}} P_x,\\
\frac{d P_y}{dt}= P_z \omega_x (t) - P_x \omega_z (t) -\frac{1}{T_{2,0}} P_y,\\
\frac{d P_z}{dt}= P_x \omega_y (t) - P_y \omega_x (t) -\frac{1}{T_{1,0}} P_z + \gamma_{\textrm{se}}(P_{\textrm{Rb}}-P_z),
\end{cases}
\end{equation}
%\end{eqnarray}
%\end{widetext}
where $\vec{\mathbf{\omega}}=2\pi \gamma \vec{\mathbf{B}}=2\pi \gamma [B_0 + B_{\textrm{ac}}\textrm{cos}(2\pi \nu_{\textrm{ac}}t)]\hat{z}+ 2\pi \gamma B_{\textrm{f}}(t)\hat{y}$,
$\gamma\approx -1.18 \times 10^7$~Hz/T is the gyromagnetic ratio of $^{129}$Xe spins.
The feedback field is $B_{\textrm{f}}(t)=\chi P_x(t)$, where $\chi$ is the feedback gain determined by the feedback circuit.
The gain is $\chi=0$ when the feedback circuit is disabled.
$P_{\textrm{Rb}}$ is the polarization of the rubidium atoms, which depends on the optical pumping and spin-relaxation rates\cite{Walker1997}.
$\gamma_{\textrm{se}}$ is the spin-exchange rate between the $^{129}$Xe and $^{87}$Rb.
$\gamma_{\textrm{se}}(P_{\textrm{Rb}}-P_z)$ represents the spin-exchange pumping of $^{129}$Xe spins.
%The feedback field produces a torque on the spins that change the magnetization.
We now consider the spin dynamics in two cases.

(1) We consider the condition: $P_{\textrm{Rb}}>0$ (in this case, bias field is along $+z$) and $B_{\textrm{ac}}=0$.
In the beginning, the $^{129}$Xe spin polarization reaches an equilibrium state.
After that, a pulse rotates the spins with angle $\theta_0$ along the $x$,
and the evolution of the spin polarization $\vec{\mathbf{P}}$ becomes\cite{Asahi2000, Suefke2017}
\begin{equation}
\begin{array}{ccc}
P_{+} (t) &=& P_0 T_{\textrm{d}} \frac{q}{T_{2,0}} \textrm{sech} [\frac{q}{T_{2,0}}(t-t_0)] e^{i (\pi/2-2\pi \nu t)},\\
P_{z} (t) &=& P_0 T_{\textrm{d}}/ T_{2,0} \{q \textrm{tanh} [\frac{q}{T_{2,0}}(t-t_0)]-1 \},
\end{array}
\label{pxy}
\end{equation}
where $P_{+}=P_x+i P_y$, $\nu$ is the oscillation frequency, and
\begin{equation}
\begin{array}{ccc}
q &=& [1+(T_{2,0}/T_{\textrm{d}})^2 + 2 \cos \theta_0 (T_{2,0}/T_{\textrm{d}})]^{1/2},\\
t_0 &=& -\frac{T_{2,0}}{q} \textrm{tanh}^{-1} [\frac{1}{q} (\frac{T_{2,0}}{T_{\textrm{d}}} \cos \theta_0 +1)].
\end{array}
\label{qq}
\end{equation}
As demonstrated in Supplemental Materials (Section 1),
the oscillation frequency has a small shift from the Larmor frequency $\nu_0$,
i.e., $\nu=\nu_0+\Delta \nu$, arising from an effect known as frequency pulling\cite{Asahi2000}.
The shift $\Delta \nu$ linearly depends on the damping rate,
i.e., $\Delta \nu=\alpha \cdot 1/T_{\textrm{d}}$, where $\alpha=0.235$ in our experiments.
%The frequency shift induced by radiation damping arises from an effect known as frequency pulling (see Supplementary Note 1 for more discussions).}
Note that longitudinal relaxation is neglected here.
$|P_{+} (t)|$ reaches its maximum value at $t=t_0$.
We discuss two cases for the initial angle $\theta_0$:
(i) When the transverse excitation of nuclear spins is small ($|\theta_0 |\ll 1$), $q \approx (1+T_{2,0}/T_{\textrm{d}})$
and $t_0\rightarrow -\infty$. Because of $t_0 <0$, $|P_{\pm}(t)|$ should be monotonically decreasing.
Finally, we have $P_{+} (t)\approx P_0 \sin \theta_0 e^{-(1/T_{2,0}+1/T_{\textrm{d}})t} e^{i (\pi/2-2\pi \nu t)}$,
which is a single-exponential decay (see Fig.~\ref{figure-1}b);
(ii) When the initial angle is $\theta_0 \approx \pi$, $\cos \theta_0\approx -1$ and $q\approx |1-T_{2,0}/T_{\textrm{d}}|$.
In this case, the initial state of nuclear spins is prepared as the inversion of population.
Although the form of equation~(\ref{pxy}) is complex,
we can still gain information from the equation~(\ref{pxy}).
Based on the definition of $t_0$ as described above, when $T_{\textrm{d}}/T_{2,0} \leq 1$, we have $t_0 \leq 0$ and thus $|P_{+}(t)|$ should be monotonically decreasing;
when $T_{\textrm{d}}/T_{2,0} < 1$, we have $t_0 > 0$ and $|P_{+}(t)|$ increases to be maximum at $t=t_0$ and then decreases to be zero (see Fig.~\ref{figure-1}c).
%{\color{red}When the maser threshold is satisfied, the magnitude of nuclear magnetization can be amplified at $t=t_0$ compared with the initial magnitude.}

(2) We consider the condition: $P_{\textrm{Rb}}<0$ (in this case, bias field is along $-z$) and $T_{\textrm{d}}/T_{2,0} < 1$.
The $^{129}$Xe spins are continuously pumped into the state with inverted population.
%We achieve the maser threshold: negative spin population and $T_{\textrm{rd}}/T_{2,0} < 1$.
Further, when the threshold ($T_{\textrm{d}}/T_{2,0} < 1$) is satisfied,
the damping-induced torque provides a sufficient strength of feedback field for sustaining the maser oscillation.
Based on numerical simulations,
we show that small transverse polarization component caused by misalignment or quantum fluctuation is sufficient for activating the Floquet maser (see Section 2 in Supplemental Materials).

~\

\noindent
\textbf{Estimation of search sensitivity for axion dark matter}

\noindent
%Ultralight bosons such as axions, axion-like particles, or dark photons are well-motivated dark matter candidates.
Here we discuss how to search for axions and axion-like particles (these are well-motivated dark matter candidates,
we call these `axions' for brevity) via our maser technique and estimate the search sensitivity.
As the nuclear spins of the maser on Earth move through the galactic dark-matter halo,
they couple to axion dark matter producing an effective oscillating magnetic field\cite{Garcon2019, budker2014},
generating axion-driven maser sidebands.
%The dark-matter-driven frequency is the Compton frequency of axions
The effective magnetic field $B_{\textrm{axion}}$:
\begin{equation}
\begin{cases}
\nu_{\textrm{axion}}=m_{\textrm{axion}}c^2/h,\\
B_{\textrm{axion}} [\textrm{T}]  \approx 6 \cdot 10^{-8} \mathrm{g}_\textrm{aNN} [\textrm{Gev}^{-1}] /\mathrm{g}_n,
\end{cases}
\label{axion}
\end{equation}
where $m_{\textrm{axion}}$ is axion mass, $c$ is the velocity of light and $h$ is the Planck constant,
$\mathrm{g}_n\approx -3/2$ is the nuclear Land$\acute{\textrm{e}}$ g-factor for $^{129}$Xe,
$\mathrm{g}_\textrm{aNN}$ is the coupling constant to be measured
that represents the coupling strength of neutrons (from $^{129}$Xe nucleus) and axions.
For example, the axion mass of $ 10^{-18}$ - $10^{-15}$~eV corresponds to the frequency $\nu_{\textrm{axion}}$ of $\sim 1$ - 100~mHz.
The experimental sensitivity (see Fig.~\ref{figure-4}c, $\delta B_{\textrm{ac}} \approx  7.2 \times 10^{-3} \nu_{\textrm{ac}} ~\textrm{nT}/\textrm{Hz}^{1/2}$) of the maser-based magnetometer to real magnetic fields can directly translate to the sensitivity to the coupling constant $\mathrm{g}_{\textrm{aNN}}$.
As a result, we achieve the search sensitivity of axions $|\mathrm{g}_\textrm{aNN} | \approx 2.7 \cdot 10^{-5} \nu_{\textrm{axion}} ~\textrm{Gev}^{-1}/\textrm{Hz}^{1/2}$,
with a higher sensitivity with smaller $\nu_{\textrm{axion}}$.
Benefitting from the narrow linewidth, the maser allows to detect axion-driven frequency as low as millihertz
(corresponding to axion mass $\sim 10^{-18}$~eV).
Current experiments, with measurement time $T_\textrm{m}=10^4$~s, $\nu_{\textrm{DM}}=1$~mHz,
yield a potential of limit on $|\mathrm{g}_\textrm{aNN} | \approx 2.7\cdot 10^{-10}~\textrm{Gev}^{-1}$,
well beyond the most stringent existing constraints\cite{Wu2019, Garcon2019}.

~\

\noindent
\textbf{{\color{red}SUPPLEMENTARY MATERIALS}}

\noindent
Supplementary Materials and Methods\\
Section 1. Damping-induced frequency shift.\\
Section 2. Numerical simulation of Floquet-maser dynamics.\\
Figure S1. Damping-induced frequency shift.\\
Figure S2. Simulation of $^{129}$Xe Floquet maser activation with different initial polarizations.\\

%\section*{REFERENCES AND NOTES}
\bibliography{scibib}

\begin{thebibliography}{99}

%maser
  % \bibitem{Townes1954} Gordon, J. P, Zeiger, H. J. $\&$ Townes, C. H. Molecular microwave oscillator and new hyperfine structure in the microwave spectrum of NH$_3$. \emph{Phys. Rev.} \textbf{95}, 282 (1954).
   \bibitem{Townes1955} J. P. Gordon, H. J. Zeiger, C. H. Townes, The maser$-$new type of microwave amplifier, frequency standard, and spectrometer. \emph{Phys. Rev.} \textbf{99}, 1264 (1955).
   \bibitem{Oxborrow2012} M. Oxborrow, J. D. Breeze, N. M. Alford, Room-temperature solid-state maser. \emph{Nature} \textbf{488}, 353 (2012).
   \bibitem{Kraus2014} H. Kraus, V. A. Soltamov, D. Riedel, S. V$\ddot{\textrm{a}}$th, F. Fuchs, A. Sperlich, P. G. Baranov, V. Dyakonov, G. V. Astakhov, Room-temperature quantum microwave emitters based on spin defects in silicon carbide. \emph{Nat. Phys.} \textbf{10}, 157 (2014).
   \bibitem{Breeze2018} J. D. Breeze, E. Salvadori, J. Sathian, N. M. Alford, C. W. Kay, Continuous-wave room-temperature diamond maser. \emph{Nature} \textbf{555}, 493 (2018).

   %laser
  % \bibitem{Schawlow1958} Schawlow, A. L. $\&$ Townes, C. H. Infrared and optical masers. \emph{Phys. Rev.} \textbf{112}, 1940 (1958).
   \bibitem{Maiman1960} T. Maiman, Stimulated optical radiation in ruby. \emph{Nature} \textbf{187}, 493 (1960).

   \bibitem{Goldenberg1960} H. M. Goldenberg, D. Kleppner, N. F. Ramsey, Atomic hydrogen maser. \emph{Phys. Rev. Lett.} \textbf{5}, 361 (1960).

   \bibitem{Cook1961} J. J. Cook, L.G. Cross, M. E., Bair, R. W. Terhune, A low-noise X-band radiometer using maser. \emph{Proceedings of the IRE}, \textbf{49}, 768 (1961).

   \bibitem{Chu2004} K. R. Chu, The electron cyclotron maser. \emph{Rev. Mod. Phys.} \textbf{76}, 489 (2004).

   \bibitem{Suefke2017} M. Suefke, S. Lehmkuhl, A. Liebisch, B. Bl$\ddot{\textrm{u}}$mich, S. Appelt, Para-hydrogen raser delivers sub-millihertz resolution in nuclear magnetic resonance. \emph{Nat. Phys.} \textbf{13}, 568 (2017).


   \bibitem{Gilles2003} H. Gilles, Y. Monfort, J. Hamel, $^3$He maser for earth magnetic field measurement. \emph{Rev. Sci. Instrum.} \textbf{74}, 4515 (2003).

   \bibitem{Bevington2020} P. Bevington, R. Gartman, W. Chalupczak, Magnetic induction tomography of structural defects with alkali¨Cmetal spin maser. \emph{Appl. Opt.} \textbf{59}, 2276 (2020).

    \bibitem{Jin2015} L. Jin, M. Pfender, N. Aslam, P. Neumann, S. Yang, J. Wrachtrup, R. B. Liu, Proposal for a room-temperature diamond maser. \emph{Nat. Commun.} \textbf{6}, 8251 (2015).

%fundamental application of masers
   \bibitem{Bear2000} D. Bear, R. E. Stoner, R. L. Walsworth, V. A. Kosteleck$\acute{\textrm{y}}$, C. D. Lane, Limit on Lorentz and CPT violation of the neutron using a two-species noble-gas maser. \emph{Phys. Rev. Lett.} \textbf{85}, 5038 (2000).

  % \bibitem{Walsworth1990} R. L. Walsworth, I. F. Silvera, E. M. Mattison, R. F. Vessot, Test of the linearity of quantum mechanics in an atomic system with a hydrogen maser. \emph{Phys. Rev. Lett.} \textbf{64}, 2599 (1990).

   \bibitem{Derevianko2014} A. Derevianko, M. Pospelov, Hunting for topological dark matter with atomic clocks. \emph{Nat. Phys.} \textbf{10}, 933 (2014).

   %He and Xe maser
 %  \bibitem{Asahi2000} T. Sato, Y. Ichikawa, S. Kojima, C. Funayama, S. Tanaka, T. Inoue, A. Uchiyama, A. Gladkov, A. Takamine, Y. Sakamoto, Y. Ohtomo, C. Hirao, M. Chikamori, E. Hikota, T. Suzuki, M. Tsuchiya,
 %  T. Furukawa, A. Yoshimi, C. P. Bidinosti, T. Ino, H. Ueno, Y. Matsuo, T. Fukuyama, N. Yoshinaga, Y. Sakemi, K. Asahi, Development of co-located $^{129}$Xe and $^{131}$Xe nuclear spin masers with external feedback scheme. % \emph{Phys. Lett. A} \textbf{382}, 588 (2018).

   \bibitem{Asahi2000} T. Sato \emph{et al.}, Development of co-located $^{129}$Xe and $^{131}$Xe nuclear spin masers with external feedback scheme. \emph{Phys. Lett. A} \textbf{382}, 588 (2018).


\bibitem{Moessner2017} R. Moessner, S. L. Sondhi, Equilibration and order in quantum Floquet matter. \emph{Nat. Phys.} \textbf{13}, 424 (2017).

%sub-millihertz
   \bibitem{Mateos2015} I. Mateos, B. Patton, E. Zhivun, D. Budker, D. Wurm, J. Ramos$-$Castro, Noise characterization of an atomic magnetometer at sub-millihertz frequencies. \emph{Sens. and Actuators A} \textbf{224}, 147 (2015).
   \bibitem{Marfaing2009} J. Marfaing \emph{et al.}, About the world-wide magnetic-background noise in the millihertz frequency range. \emph{Europhys. Lett.} \textbf{88}, 19002 (2009).


%EDM and axion
  % \bibitem{graham2013} P. W. Graham, S. Rajendran, New observables for direct detection of axion dark matter. \emph{Phys. Rev. D} \textbf{88}, 035023 (2013).
  \bibitem{DeMille2017} D. DeMille, J. M. Doyle, A. O. Sushkov, Probing the frontiers of particle physics with tabletop-scale experiments. \emph{Science} \textbf{357}, 990 (2017).

  \bibitem{Safronova2018} M. S. Safronova, D. Budker, D. DeMille, Derek F. Jackson Kimball, A. Derevianko, C. W. Clark, Search for New Physics with Atoms and Molecules. \emph{Rev. Mod. Phys.} \textbf{90}, 025008 (2018).


 %  \bibitem{Graham2018} P. W. Graham \emph{et al.}, Spin precession experiments for light axionic dark matter. \emph{Phys. Rev. D} \textbf{97}, 055006 (2018).


%Floquet matter

   \bibitem{Zhang2017} J. Zhang, P. W. Hess, A. Kyprianidis, P. Becker, A. Lee, J. Smith, G. Pagano, I.-D. Potirniche, A. C. Potter, A. Vishwanath, N. Y. Yao, C. Monroe, Observation of a discrete time crystal. \emph{Nature} \textbf{543}, 217 (2017).

   \bibitem{Rechtsman2013} M. C. Rechtsman, J. M. Zeuner, Y. Plotnik, Y. Lumer, D. Podolsky, F. Dreisow, S. Nolte, M. Segev, A. Szameit, Photonic Floquet topological insulators. \emph{Nature} \textbf{496}, 196 (2013).


   \bibitem{Eckardt2017} Eckardt, A. Colloquium: Atomic quantum gases in periodically driven optical lattices. \emph{Rev. Mod. Phys.} \textbf{89}, 011004 (2017).

   \bibitem{Shu2018} Z. Shu, Y. Liu, Q. Cao, P. Yang, S. Zhang, M. B. Plenio, F. Jelezko, J. Cai, Observation of Floquet Raman Transition in a Driven Solid-State Spin System. \emph{Phys. Rev. Lett.} \textbf{121}, 210501 (2018).

%Floquet spectroscopy application
   \bibitem{Lang2015} J. E. Lang, R. B. Liu, T. S. Monteiro, Dynamical-decoupling-based quantum sensing: Floquet spectroscopy. \emph{Phys. Rev. X} \textbf{5}, 041016 (2015).
   \bibitem{Joas2017} T. Joas, A. M. Waeber, G. Braunbeck, F. Reinhard, Quantum sensing of weak radio-frequency signals by pulsed Mollow absorption spectroscopy. \emph{Nat. Commun.} \textbf{8}, 964 (2017).
   \bibitem{Zhang1994} Y. Zhang, M. Ciocca, L. W. He, C. E. Burkhardt, J. J. Leventhal, Measurement of atomic polarizabilities using Floquet spectroscopy. \emph{Phys. Rev. A} \textbf{50}, 1101 (1994).

%axion
\bibitem{budker2014} D. Budker, P. W. Graham, M. Ledbetter,S. Rajendran, A. O. Sushkov, Proposal for a cosmic axion spin precession experiment (CASPEr). \emph{Phys. Rev. X} \textbf{4}, 021030 (2014).
   \bibitem{Wu2019} T. Wu \emph{et al.}, Search for axionlike dark matter with a liquid-state nuclear spin comagnetometer. \emph{Phys. Rev. Lett.} \textbf{122}, 191302 (2019).
   \bibitem{Garcon2019} A. Garcon \emph{et al.}, Constraints on bosonic dark matter from ultralow-field nuclear magnetic resonance. \emph{Sci. Adv.} \textbf{5}, eaax4539 (2019).

%nonlinear multiphoton
 \bibitem{Glasenapp2014} P. Glasenapp, N. A. Sinitsyn, Luyi Yang, D. G. Rickel, D. Roy, A. Greilich, M. Bayer, and S. A. Crooker, Spin noise spectroscopy beyond thermal equilibrium and linear response. \emph{Phys. Rev. Lett.} \textbf{113}, 156601 (2014).

%magnetometer
   \bibitem{Greenberg1998} Y. S. Greenberg, Application of superconducting quantum interference devices to nuclear magnetic resonance. \emph{Rev. Mod. Phys.} \textbf{70}, 175 (1998).
   \bibitem{DBudker2007} D. Budker,  M. V. Romalis, Optical magnetometry. \emph{Nat. Phys.} \textbf{3}, 227 (2007).
   \bibitem{Taylor2008} J. M. Taylor \emph{et al.}, High-sensitivity diamond magnetometer with nanoscale resolution. \emph{Nat. Phys.} \textbf{4}, 810 (2008).

 %  \bibitem{Kominis2003} I. K. Kominis, T. W. Kornack, J. C. Allred, M. V. Romalis, A subfemtotesla multichannel atomic magnetometer. \emph{Nature} \textbf{422}, 596 (2003).


%sideband detection
 %  \bibitem{Dagenais1988} D. M. Dagenais, F. Bucholtz, K. P. Koo, Elimination of residual signals and reduction of noise in a low-frequency magnetic fiber sensor. \emph{Appl. Phys. Lett.} \textbf{53}, 1474 (1988).
 %  \bibitem{Garcon2017} A. Garcon \emph{et al.}, The Cosmic Axion Spin Precession Experiment (CASPEr): a dark-matter search with nuclear magnetic resonance. \emph{Quant. Sci. Technol.} \textbf{3}, 014008 (2017).



%SEOP and detection
   \bibitem{Walker1997} T. G. Walker, W. Happer, Spin-exchange optical pumping of noble$-$gas nuclei. \emph{Rev. Mod. Phys.} \textbf{69}, 629 (1997).
 %  \bibitem{Gentile2017} T. R. Gentile, P. J. Nacher, B. Saam, T. G. Walker, Optically polarized $^3$He. \emph{Rev. Mod. Phys.} \textbf{89}, 045004 (2017).



%   \bibitem{Chupp1994} T. E. Chupp, R. J. Hoare, R. L. Walsworth, B. Wu, Spin-exchange-pumped $^3$He and $^{129}$Xe Zeeman masers. \emph{Phys. Rev. Lett.} \textbf{72}, 2363 (1994). %He3 and Xe129
 %  \bibitem{Yoshimi2002} A. Yoshimi \emph{et al.}, Nuclear spin maser with an artificial feedback mechanism. \emph{Phys. Lett. A} \textbf{304}, 13 (2002).


   \bibitem{Bienfait2016} A. Bienfait \emph{et al.}, Controlling spin relaxation with a cavity. \emph{Nature} \textbf{531}, 74 (2016).

  %radiation damping
  \bibitem{Bloembergen1954} N. Bloembergen, R. V. Pound, Radiation damping in magnetic resonance experiments. \emph{Phys. Rev.} \textbf{95}, 8 (1954).

%  \bibitem{Meiboom1959} A. Sz$\ddot{\textrm{o}}$ke, S. Meiboom, Radiation damping in nuclear magnetic resonance. \emph{Phys. Rev.} \textbf{113}, 585 (1959).
%  \bibitem{Bloom1957} S. Bloom, Effects of radiation damping on spin dynamics. \emph{J. Appl. Phys.} \textbf{28}, 800 (1957).

  \bibitem{Shirley1965} J. H. Shirley, Solution of the Schr$\ddot{\textrm{o}}$dinger equation with a Hamiltonian periodic in time. \emph{Phys. Rev.} \textbf{138}, B979 (1965).
  \bibitem{Cohen-Tannoudji1992} C. Cohen-Tannoudji, J. Dupont-Roc, G. Grynberg, \emph{Atom-Photon Interactions: Basic Processes and Applications} (John Wiley $\&$ Sons, New York, 1992).

  \bibitem{Gentile2017} T. R. Gentile, P. J. Nacher, B. Saam, T. G. Walker, Optically polarized $^3$He. \emph{Rev. Mod. Phys.} \textbf{89}, 045004 (2017).

  \bibitem{Benedict2018} M. G. Benedict, \emph{Super-radiance: Multiatomic coherent emission.} Routledge (2018).

%strong coupling
  \bibitem{Fuchs2009} G. D. Fuchs, V. V. Dobrovitski, D. M. Toyli, F. J. Heremans, D. D. Awschalom, Gigahertz dynamics of a strongly driven single quantum spin. \emph{Science} \textbf{326}, 1520 (2009).

%   \bibitem{Lewenstein1990} M. Lewenstein, Y. Zhu, T. W. Mossberg, Two-photon gain and lasing in strongly driven two-level atoms. \emph{Phys. Rev. Lett.} \textbf{64}, 3131 (1990).

  \bibitem{Kominis2003} I. K. Kominis, T. W. Kornack, J. C. Allred, M. V. Romalis, A subfemtotesla multichannel atomic magnetometer. \emph{Nature} \textbf{422}, 596 (2003).

  % \bibitem{Ernst1987} Ernst, R. R., Bodenhausen, G. $\&$ Wokaun, A. \emph{Principles of Nuclear Magnetic Resonance in One and Two Dimensions} (Clarendon Press, Oxford, 1987).

 \bibitem{Tannoudji1996} C. Cohen$-$Tannoudji, The Autler-Townes effect revisited. In: Amaz. Light, Springer \textbf{109} (1996).




 %  \bibitem{Chen2011} Chen, H. Y., Lee, Y., Bowen, S. $\&$ Hilty, C. Spontaneous emission of NMR signals in hyperpolarized proton spin systems. \emph{J. Magn. Reson.} \textbf{208}, 204 (2011).




% \bibitem{Kimball2017} D. F. Kimball \emph{et al.}, Overview of the cosmic axion spin precession experiment (CASPEr). \emph{arXiv}:1711.08999. (2017).


%   \bibitem{LAllen1987} Allen, L. $\&$ Eberly, J. H. \emph{Optical Resonance and Two-Level Atoms} (Dover Publications, New York, 1987).

    %NMR and NMR QIP

  % \bibitem{Vandersypen2005} Vandersypen, L. M. K. $\&$ Chuang, I. L. NMR techniques for quantum control and computation. \emph{Rev. Mod. Phys.} \textbf{76}, 1037 (2005).

   %time crystal
  % \bibitem{Choi2017} Choi, S. \emph{et al.} Observation of discrete time-crystalline order in a disordered dipolar many-body system. \emph{Nature} \textbf{543}, 221 (2017).

  % \bibitem{Moessner2017} Moessner, R. $\&$ Sondhi, S. L. Equilibration and order in quantum Floquet matter. \emph{Nat. Phys.} \textbf{13}, 424 (2017).

  % \bibitem{Zhang2017} Zhang, J. \emph{et al.} Observation of a discrete time crystal. \emph{Nature} \textbf{543}, 217 (2017).

   %spin noise
  % \bibitem{Glasenapp2014} Glasenapp, P. \emph{et al.} Spin noise spectroscopy beyond thermal equilibrium and linear response. \emph{Phys. Rev. Lett.} \textbf{113}, 156601 (2014).

  % \bibitem{Li2016} Li, J., Lu, D., Luo, Z., Laflamme, R., Peng, X. $\&$ Du, J. Approximation of reachable sets for coherently controlled open quantum systems: Application to quantum state engineering. \emph{Phys. Rev. A} \textbf{94}, 012312 (2016).


   %decoherence limit
 %  \bibitem{Kusch1956} P. Kusch, Phys. Rev. \textbf{101}, 627 (1956).



  % \bibitem{Beige2000} Beige, A., Braun, D., Tregenna, B. $\&$ Knight, P. L. Quantum computing using dissipation to remain in a decoherence-free subspace. \emph{Phys. Rev. Lett.} \textbf{85}, 1762 (2000).




 %  \bibitem{Vijay2012} Vijay, R. \emph{et al.} Stabilizing Rabi oscillations in a superconducting qubit using quantum feedback. \emph{Nature} \textbf{490}, 77 (2012).

   %self-oscillating magnetometers
 %  \bibitem{Higbie2006} Higbie, J. M., Corsini, E. $\&$ Budker, D. Robust, high-speed, all-optical atomic magnetometer. \emph{Rev. Sci. Instr.} \textbf{77}, 113106 (2006).

 %  \bibitem{Wilson2015} Wilson, D. J., Sudhir, V., Piro, N., Schilling, R., Ghadimi, A. $\&$ Kippenberg, T. J. Measurement-based control of a mechanical oscillator at its thermal decoherence rate. \emph{Nature} \textbf{524}, 325 (2015).

   %feedback mechanism
 %  \bibitem{Geremia2003} Geremia, J. M., Stockton, J. K., Doherty, A. C. $\&$ Mabuchi, H. Quantum Kalman filtering and the Heisenberg limit in atomic magnetometry. \emph{Phys. Rev. Lett.} \textbf{91}, 250801 (2003).
 %  \bibitem{Behbood2013} Behbood, N., Colangelo, G., Ciurana, F. M., Napolitano, M., Sewell, R. J. $\&$ Mitchell, M. W. Feedback cooling of an atomic spin ensemble. \emph{Phys. Rev. Lett.} \textbf{111}, 103601 (2013).

     %SI
  % \bibitem{SI} See Supplementary note for details of the calculations and experimental procedures.

%   \bibitem{Waker1989} T. G. Walker, Phys. Rev. A \textbf{40}, 4959 (1989).



 % \bibitem{Vysotsky1978} M. Vysotsky, Y. Zeldovich, M. Khlopov, V. Chechetkin, Some astrophysical limitations on the axion mass. \emph{JETP Lett.} \textbf{27}, 533 (1978).


  \bibitem{Meschede1985} D. Meschede, H. Walther, G. M$\ddot{\textrm{u}}$ller, One-atom maser. \emph{Phys. Rev. Lett.} \textbf{54}, 551 (1985).


  % \bibitem{Glenday2008} A. G. Glenday, C. E. Cramer, D. F. Phillips, R. L. Walsworth, Limits on anomalous spin-spin couplings between neutrons. \emph{Phys. Rev. Lett.} \textbf{101}, 261801 (2008).




  % \bibitem{Russomanno2017} Russomanno, A. $\&$ Santoro, G. E. Floquet resonances close to the adiabatic limit and the effect of dissipation. \emph{J. Stat. Mech.} \textbf{2017}, 103104 (2017).
  % \bibitem{Sias2008} Sias, C. \emph{et al.} Observation of photon-assisted tunneling in optical lattices. \emph{Phys. Rev. Lett.} \textbf{100}, 040404 (2008).
  % \bibitem{Shu2018} Shu, Z. \emph{et al.} Observation of Floquet Raman transition in a driven solid-state spin system. \emph{Phys. Rev. Lett.} \textbf{121}, 210501 (2018).



%\bibitem{Bulatowicz2013} Bulatowicz, M. \emph{et al.} Laboratory Search for a Long-Range T-Odd, P-Odd Interaction from Axionlike Particles Using Dual-Species Nuclear Magnetic Resonance with Polarized $^{129}$Xe and $^{131}$Xe Gas. \emph{Phys. Rev. Lett.} \textbf{111}, 102001 (2013).
  % \bibitem{Korver2015} Korver, A. \emph{et al.} Synchronous spin$-$exchange optical pumping. \emph{Phys. Rev. Lett.} \textbf{115}, 253001 (2015).
 %  \bibitem{Limes2018} Limes, M. E., Sheng, D. $\&$ Romalis, M. V. $^3$He$-$$^{129}$Xe Comagnetometery using $^{87}$Rb Detection and Decoupling. \emph{Phys. Rev. Lett.} \textbf{120}, 033401 (2018).

%  \bibitem{Ma2011} Z. L. Ma, E. G. Sorte, B. Saam, Collisional $^3$He and $^{129}$Xe Frequency Shifts in Rb-Noble-Gas Mixtures. \emph{Phys. Rev. Lett.} \textbf{106}, 193005 (2011).


%   \bibitem{Raffelt2008} G. Raffelt, Astrophysical Axion Bounds (Springer, 2008).


%ZULF
%   \bibitem{Ledbetter2011} M. P. Ledbetter \emph{et al.}, Near-zero-field nuclear magnetic resonance. \emph{Phys. Rev. Lett.} \textbf{107}, 107601 (2011).
 %  \bibitem{Jiang2018} M. Jiang, T. Wu, J. W. Blanchard, G. Feng, X. Peng, D. Budker, Experimental benchmarking of quantum control in zero-field nuclear magnetic resonance. \emph{Sci. adv.} \textbf{4}, eaar6327 (2018).


\end{thebibliography}

\bibliographystyle{Science}

~\

\noindent
\textbf{Acknowledgments}: We thank D. Suter, D. Sheng, W. H. Hai, M. Gu, Y. S. Zhang, J. M. Cai, A. Garcon, K. Asahi, T. E. Chupp, R. L. Walsworth, and J. W. Blanchard for valuable discussions.
\textbf{Funding}: This work was supported by National Key Research and Development Program of China (Grant No. 2018YFA0306600), National Natural Science Foundation of China (Grants Nos. 11425523, 11661161018), Anhui Initiative in Quantum Information Technologies (Grant No. AHY050000), National Science Foundation (Grant ECCS 1710558). $\textrm{D. B.}$ acknowledges the support of the European Research Council under the European Union's Horizon 2020 Research and Innovation Program under Grant agreement No. 695405 and by the DFG under the Reinhart Koselleck program and by the Cluster of Excellence ¡°Precision Physics, Fundamental Interactions, and Structure of Matter¡± ($\textrm{PRISMA}^{+}$ EXC 2118/1) funded by the German Research Foundation (DFG) within the German Excellence Strategy (Project ID 39083149).
\textbf{Author contributions}:
M. J. designed and performed experiments, analyzed the data and wrote the manuscript.
H. W. S. and Z. W. performed the measurements and analyzed the data.
X. H. P. proposed the experimental concept, devised the experimental protocols, and wrote the manuscript.
D. B. contributed to the design of the experiment and proofread and edited the manuscript.
All authors contributed with discussions and checking the manuscript.

% For your review copy (i.e., the file you initially send in for
% evaluation), you can use the {figure} environment and the
% \includegraphics command to stream your figures into the text, placing
% all figures at the end.  For the final, revised manuscript for
% acceptance and production, however, PostScript or other graphics
% should not be streamed into your compliled file.  Instead, set
% captions as simple paragraphs (with a \noindent tag), setting them
% off from the rest of the text with a \clearpage as shown  below, and
% submit figures as separate files according to the Art Department's
% instructions.

\clearpage

%fig1
%\newpage

%fig2
%\newpage

%fig3
%\newpage

%fig5

\end{document}